\documentclass{ws-ijmpa}

\begin{document}

\markboth{Li and Zhu}
{QCD Effects in High Energy Processes}

\title{QCD Effects in High Energy Processes}
\author{Chong Sheng Li and Shou-hua Zhu}
\address{Department of Physics, Peking University, Beijing 100871, China}

\maketitle

\begin{abstract}
  In this talk, some important QCD effects in Higgs physics, supersymmetry and top
  physics, as well as the factorization and resummation techniques in QCD are reviewed.
\end{abstract}

\keywords{QCD; Higgs; supersymmetry; top; factorization and resummation.}

\section{QCD effects in Higgs physics}

The Higgs boson is essential in the Standard Model (SM) for explaining the electroweak
symmetry breaking and the origin of the mass of the fermions. Beyond the SM, we may also
have different variations of the SM Higgs boson,viz. Two Higgs doublet model(2HDM)
\cite{THDM}, 3HDM \cite{3HDM}, triplet \cite{Triplet}, SUSY \cite{mssm}, little
\cite{lhiggs} and fat \cite{fath} Higgs bosons etc. However none of them have yet been
observed. Thus the search for Higgs bosons becomes a primary goal of next generation
colliders.

Before the Higgs is found, we may ask some questions like: (1) Is there any fundamental
scalar field in nature? (2)What kind of Higgs might be found? All these questions may be
answered at the next generation colliders. And QCD effects play an important role in
answering the above questions because they will affect Higgs production and decay at next
generation colliders.

\subsection{The SM Higgs at Hadron Colliders}

{\em $gg\rightarrow H$:} Gluon fusion is the dominant production mode for the Higgs boson
at hadron colliders. The NLO QCD corrections were calculated years ago and found to be
large\cite{nggh}. Since then the NNLO QCD corrections were done, which contains: (1)
Two-loop corrections to the H-g-g vertex in Ref.~\cite{nnlo1}, (2) Soft-plus-virtual
gluon corrections in Ref.~\cite{nnlo2}, and (3) Two-to-three body processes contributions
in Ref.~\cite{nnlo3}. The calculations show that the NNLO corrections significantly
reduced the renormalization and factoriaztion scales dependence.

For this production channel, for $100$GeV $\leq m_H \leq$ $140$GeV, $H\rightarrow
\gamma\gamma$ is the most important decay mode for searching for the Higgs boson. To
thoroughly analyze the signal and the background, $gg\rightarrow\gamma\gamma$ has been
calculated at the NLO level in Ref.~\cite{ggrr}.

Also, the effect of the jet veto on $gg\rightarrow H$ has been studied at the NLO level
in Ref.~\cite{jetv}. The jet veto reduces both the total cross section and the size of
the higher-order QCD corrections.

{\em $pp(\bar{p})\rightarrow t\bar{t}H, b\bar{b}H$:} Although rates for these production
channels are small, the final states have distinctive signatures. At the LHC, the
$t\bar{t}H$ channel can complement WH associated production for $m_H\leq 130$GeV. Beyond
the SM, the $b\bar{b}H$ channel may be more important since the $b\bar{b}H$ coupling can
be enhanced, for example, in the MSSM the vertex can be enhanced by $\tan\beta$.

The NLO QCD corrections to $pp(\bar{p})\rightarrow t\bar{t}H$ have been calculated in
Ref.~\cite{htt}, which increase the LO total cross sections and significantly reduce the
scale dependence of the LO results. The NLO QCD corrections to $pp(\bar{p})\rightarrow
b\bar{b}H$ have been calculated in Ref.~\cite{hbb,hj1}. Moreover, in the MSSM, the
SUSY-QCD corrections to $pp(\bar{p})\rightarrow t\bar{t}H$ were also done in
Ref.~\cite{ma}, and the partly SUSY-QCD corrections to $pp(\bar{p})\rightarrow b\bar{b}H$
were examined in Ref.~\cite{gg2} in the special case for convenience that the relevant
sparticles are heavy, well above TeV scale.

{\em $b\bar{b} \rightarrow H$:} Due to the ability to tag the b quark in the final state,
the processes of the Higgs boson production associated with b quark(s) are promising. So
these production channels may be more important since the $bbH$ coupling can be enhanced
beyond the SM. For inclusive Higgs production, the LO process is $b\bar{b}\rightarrow H$.
And in its NLO QCD calculations, due to the smallness of the b quark mass, large
logarithms $\log(Q^2/m^2_b)$ might arise from phase space integration in the
four-flavor-number scheme and the convergence of the perturbative expansion can be
improved by summing the collinear logarithms to all orders by the introduction of evolved
b quark parton distributions with an appropriate factorization scale, as analysized in
Ref.~\cite{will}: Since in the collinear limit $d\sigma/dt\sim 1/t$, one can observe that
the collinear limit here is about $\sqrt{-t}<m_H/4$. Thus the collinear logarithm that is
generated at NLO is therefore approximately $\ln(m_h/4\mu_F)$, rather than
$\ln(m_h/\mu_F)$, leading to the fact that the factorization scale for the
$1/\ln(m_h/m_b)$ correction should be chosen to be of order $\mu_F \approx m_h/4$ in
order to sum the collinear logarithm into the parton distribution function and thus the
results obtained in the bottom parton picture can be in good agreement with those for the
gluon-initiated production process. However there are still many open questions
concerning the introducing of bottom densities, and due to the limited space, for more
discussions please see for example Ref.~\cite{hj2}.

Moreover, if at least one high-$p_T$ b quark is required to be observed, the leading
partonic process is $gb\rightarrow bH$~\cite{Huang:1998vu}. The relevant NLO QCD
calculations were done in Ref.~\cite{will,gbjet}, and the SUSY-QCD corrections in
Ref.~\cite{sqcdgb}.

{\em $q\bar{q} \rightarrow q\bar{q}VV^* \rightarrow q\bar{q}H$:} The weak-boson fusion
process is expected to provide crucial information on Higgs boson couplings at the LHC.
The NLO QCD calculations have been done\cite{WBF}. Since then the QCD corrections to jet
correlations in this channel have been done\cite{WBFJ} and are shown to be modest, of the
order of from $5\%$ to $10\%$ in most cases, but reaching 30$\%$ occasionally, and the
scale uncertainties range from the order of 5$\%$ or less for distributions to below
±2$\%$ for the Higgs boson cross section in typical weak-boson fusion search regions.

\subsection{The SM Higgs at Linear Colliders}

{\em $e^+e^-\rightarrow t\bar{t}H$:} Although the production rate is small, it has a
distinctive signature and can potentially be used to measure the relevant Yukawa
coupling. The NLO QCD corrections were done in Ref.~\cite{eetth} and found to enhance the
total cross sections by a factor of roughly 1.5 at $\sqrt{s}=500$GeV.

{\em $\gamma\gamma\rightarrow t\bar{t}H$:} An $e^+e^-$ LC can also be designed to operate
as a $\gamma\gamma$ collider. Thus $e^+e^-\rightarrow \gamma\gamma \rightarrow t\bar{t}H$
offers another approach to probe the Higgs boson and the relevant Yukawa coupling in
addition to $e^+e^-\rightarrow t\bar{t}H$. The NLO QCD corrections were done in
Ref.~\cite{rrtth} and can reach reach $34.8\%$ at $\sqrt{s}=800$GeV.

\subsection{The MSSM Higgs at Hadron Colliders}

The Higgs sector of MSSM, which is the special case of the 2HDM, is of particular
theoretical interest, and contains five physical Higgs bosons: two neutral CP-even bosons
$h^0$ and $H^0$, one neutral CP-odd boson $A^0$, and two charged bosons $H^\pm$. The
$h^0$ is the lightest, with a mass $m_{h^0}\leq140$\, GeV when including the radiative
corrections\cite{massh0}, and is a SM-like Higgs boson especially in the decoupling
region ($m_{A^0}\gg m_{Z^0}$). The other four are non-SM-like, and the discovery of them
may give the direct evidence of MSSM. It has been shown in Ref.~\cite{detect,LHCHiggs}
that the $h^0$ boson of MSSM cannot escape detection at the CERN LHC and that more than
one neutral Higgs particle can be found in a large area of the supersymmetry (SUSY)
parameter space.

Many higher order QCD corrections to Higgs bosons production have been performed, for
example, $pp \rightarrow$ (pseudo)scalar Higgs bosons at NNLO in Ref.~\cite{susyh}. Note
that for large values of $\tan\beta$ the bottom loop contribution to $pp \rightarrow$
pseudo-scalar Higgs bosons is dominant. It is necessary to point out that the NNLO
corrections to this process have been obtained only in the infinite top mass limit and
are not applicable for large values of $\tan\beta$ so that one has to rely on the full
NLO results of the last 2 papers of Ref.~\cite{nggh}. The effects of SUSY-QCD in hadronic
Higgs production at NNLO in Ref.~\cite{har2}, and the NLO QCD corrections to Higgs plus 1
or 2 high $P_T$ final bottom quark(s) production in Ref.~\cite{hj1,hj2} and
Ref.~\cite{hbb,hj1,hj3}, respectively.

For charged Higgs production, the NLO QCD and SUSY-QCD corrections to $gb \rightarrow
tH^-$, which is the primary charged Higgs boson production channel at the LHC, were done
in Ref.~\cite{gbth} and Ref.~\cite{gbthsqcd}, respectively. The size of the QCD
corrections are quite large and can reach 80$\%$. The NLO QCD and SUSY-QCD corrections to
$b\bar{b} \rightarrow W^+H^-$ were done in Ref.~\cite{bbwh} and Ref.~\cite{zhaojun},
respectively. The ones to $b\bar{b}\rightarrow H^+H^-$ were done in Ref.~\cite{hpair}.

In Ref.~\cite{hhpair} and Ref.~\cite{jin2}, the QCD and SUSY-QCD effects on neutral Higgs
bosons pair production through $q\bar{q}$ annihilation, $gg$ fusion and $b\bar{b}$
annihilation were calculated, respectively. In Ref.~\cite{az}, the NLO QCD and SUSY-QCD
corrections to $A^0Z^0$ associated production were calculated. For a cross check, both
the dimensional regularization scheme and the dimensional reduction scheme were used to
organize the calculations which yielded the same NLO rates in Ref.~\cite{jin2,az}. The
NLO corrections can either enhance or reduce the total cross sections, but it generally
significantly reduces the dependence of the total cross sections on the
renormalization/factorization scale. The uncertainty of the total cross sections, due to
the parton distribution function uncertainties, was also studied in Ref.~\cite{jin2,az}
and it was found that the NLO QCD corrections do not reduce the PDF uncertainty.

\subsection{The MSSM Higgs at Linear Colliders}

At the linear colliders, the NLO (SUSY)QCD corrections to $e^+e^-\rightarrow
t\bar{t}H(b\bar{b})$ were done in Ref.~\cite{lctth}. The QCD corrections significantly
increase the total cross sections for $\sqrt{s}=500$GeV. The SUSY QCD effects generally
are very moderate (say 10$\%$) and under control. The NLO QCD corrections to
$e^+e^-\rightarrow t\bar{b}H^-$ were done in Ref.~\cite{lctbh}. After resumming the large
logarithmic corrections that arise in the on-mass-shell scheme of quark mass
renormalization by adopting the modified minimal-subtraction scheme, the convergence
behavior of the perturbative expansion is improved. The NLO QCD corrections lead to a
significant reduction of the theoretical uncertainties due to scheme and scale
dependences.

\section{QCD effects in SUSY}

As stated above, the MSSM is one of the most interesting new physics models beyond the
SM. It was devised to solve the hierarchy problem and has the celebrated feature of gauge
coupling unification, and can naturally provide a candidate for dark matter. The direct
search for SUSY particles is one of the primary task of the current and future colliders.
As in the case of Higgs searches, QCD effects are essential for the theoretical
evaluation of the production cross sections and decay rates. They could also be important
for distinguishing SUSY breaking scenarios. In the following we review the works on the
QCD corrections to the production cross sections of SUSY particles at hadron colliders
and linear colliders.

For QCD corrections, the most difficult cases for two particle final states are colored
particle pair production at hadron colliders. These have been done at next-to-leading
order for squark pair production, gluino pair production, and squark-gluino production
\cite{Phys.Rev.Lett.74.2905}. The top squark pair production was calculated in Ref.
\cite{Nucl.Phys.B515.3}. For processes involving one colored final state, the associated
production of gluinos and gauginos was calculated with QCD corrections and SUSY QCD
corrections \cite{berger,hep-ph/0211145}. The associated production of top squark and
chargino was accomplished \cite{lgjin} with both QCD and SUSY QCD corrections. For the
colorless final states, the NLO corrections to slepton pair production and gaugino pair
production are presented in Ref.~\cite{gaugino}. In R-parity violating supersymmetric
models, SUSY particles need not be produced in pairs. Single top squark production was
considered \cite{PL.B488.359}. The production of a top squark and a lepton was calculated
in Ref.~\cite{PL.B558.165}. The production of a single slepton was calculated in
Ref.~\cite{NP.B660.343,hep-ph/0507331}.

In general, the QCD corrections enhance the total cross sections by 10-90\%. More
importantly, the higher order corrections reduce the renormalization and factorization
scale dependence by a factor of 3 to 4, which render the results more stable and
reliable. The remaining scale dependence, typically at a level of 10-15\%, serves as an
estimate of the theoretical uncertainty \cite{hep-ph/9809259,hep-ph/0211145}.

At linear colliders, the situation is much simpler, since now only colored final states
receive QCD corrections. In fact, since the gluinos do not participate in electroweak
interactions, the only possible two-body final state at tree level is squark pair
production. The QCD and SUSY QCD corrections to this process at $e^+e^-$ colliders are
presented in Ref.~\cite{ee}. The complementary results in photon-photon collision are
calculated in Ref.~\cite{NP.B515.15}. For three-body final states, the QCD and SUSY QCD
corections to squark-squark-gluon production and quark-squark-gluino production have been
described in Ref.~\cite{qqg}. In general, at linear colliders, the QCD corrections are
positive and dominant over other ones at low colliding energy (e.g. 500-1000 GeV).

\section{QCD effects in top physics}

\subsection{Top quark decay and Single top production}

The one-loop QCD corrections to top quark decay were calculated years ago
\cite{topdecay1}. There were some recent works studying the QCD effects on the top quark
decay, for example, two loop calculation techniques~\cite{two}, polarized top quark
decay~\cite{pol}, decay distributions~\cite{dist} and SUSY-QCD effects in top quark rare
decay within a most general framework~\cite{susde}. However, we only discuss the top
quark production in the following and mostly focus on recent developments. We first
review the SM processes of single top production. At hadron colliders, single top quarks
can be produced within the SM in three different channels, and the corresponding NLO QCD
effects have been completed: the s-channel $W^*$ production \cite{t1}, the t-channel
W-exchange mode \cite{t2}, and through $t W^-$ production \cite{t3}. Later, a new NLO
calculation for fully differential single-top-quark final states also were
obtained~\cite{diff}. This calculation is performed using phase space slicing and dipole
subtraction methods, and the dipole subtraction method calculation retains the full spin
dependence of the final state particles. Recently, there have been several works
combining the decay effects on the production processes~\cite{diy}. In order to confront
theory with experimental data, where kinematical cuts are necessary in order to detect a
signal, it is crucial to accurately model event topologies of single top quark events.
Ref~\cite{diy} calculated the differential cross sections for on-shell single top quark
production. The complete NLO calculations including both the single top quark production
and decay have been done. In these calculations, the narrow width approximation was
adopted in order to link top quark production with its consequent decay and various
kinematic distributions are examined both with and without top quark decay at NLO.

Ref~\cite{eg} calculated the NLO QCD effects on single top production at an e$\gamma$
collider. Within new physics beyond the SM, some works studied single top FCNC production
at ILC~\cite{l1} and HERA~\cite{hera}, as well as hadron colliders(LHC)~\cite{np}. The
results at the LHC can be as large as a few pb, and may supply a powerful probe for the
details of the SUSY FCNC couplings. Moreover, the QCD corrections to single top quark
production induced by model-independent FCNC couplings have been investigated by two
groups~\cite{qcdfc}. The NLO results increase the experimental sensitivity to the
anomalous couplings, and vastly reduce the dependence of the total cross sections on the
renormalization/factorization scale at the Tevatron Run II, which leads to increased
confidence in predictions based on these results..

\subsection{Top Pair production}

In the SM, QCD corrections to the total cross sections for top quark pair production at
the Tevatron and LHC are known very well~\cite{nason}. Since then authors obtained: $p_T$
and y spectra; double-differential spectra; resummation at LL level and NLL
level~\cite{ptt}.

Recent developments(include the soft-gluon corrections at NNLO) can be seen from
Ref.~\cite{top1}. The state of art at present is: the soft NNLO corrections to the total
top quark cross section and top transverse momentum distributions in hadron-hadron
collisions with new soft NNNLL terms and some virtual terms, including all
soft-plus-virtual factorization and renormalization scale terms. It was found that these
new subleading corrections greatly diminish the dependence of the cross section on the
kinematics and on the factorization/renormalization scales.

During the past few years, spin correlations in top pair production have been studied at
various colliders:

1. Hadron colliders~\cite{spin2}: The NLO QCD effects on the hadronic production of
$t\bar t$ quarks in a general spin configuration have been computed and also the dilepton
angular correlation coefficients C that reflect the degree of correlation between the $t$
and $\bar t$ spins. These results for the Tevatron show that the scale and in particular
the PDF uncertainties in the prediction of the dileptonic angular distribution must be
reduced before $t\bar t$ spin correlations can be used in a meaningful way to search for
relatively small effects of new interactions that are, for example, not distinguished by
violating parity or CP invariance. And the LHC may turn top quark spin correlations into
a precision tool for the analysis of $t\bar t$ events.

2. Polarized photon colliders~\cite{spin3}: A variety of spin observables have been
calculated for the process $\gamma\gamma\rightarrow t\bar{t} X$ up to order $\alpha^2
\alpha_s$, especially the NLO QCD contributions to the fully differential cross section
with intermediate top quark pair production at a photon collider.

Moreover, Ref.~\cite{zhu} studied the NLO QCD effects in $VV\rightarrow tt$ at the ILC.
They found that QCD corrections can be quite substantial, so that they need to be taken
into account when studying $t\bar{t}$ production. Recently, a first paper on NLO QCD
corrections to $gg \rightarrow t \bar{t} g$ became available~\cite{tt3}. We also should
mention the works on the interference between production and decay at Linear
Colliders~\cite{lin}.

SUSY-QCD effects on top quark production at $e^+e^-$ and photon-photon colliders were
studied in Ref.~\cite{cslin} and Ref.~\cite{csli},respectively. Strong supersymmetric
quantum effects on top quark production as well as supersymmetric QCD parity
nonconservation in top quark pairs at the Tevatron were studied in Ref.~\cite{csli2}.
Ref.~\cite{csli3} studied the one-loop supersymmetric QCD corrections arising from
squarks and gluino to top quark pair production by gg fusion at the LHC in the MSSM, and
found that the corrections to the hadronic cross section amount to a few percent.
$O(\alpha_s)$ QCD Corrections to spin correlations in $e^- e^+ \rightarrow t \bar t$
process at the NLC and SUSY-QCD at $e^+e^-$ colliders(including spin correlations) can be
found in Ref.~\cite{spin1}.

\section{Factorization and Resummation}


The QCD factorization theorems \cite{hep-ph/0409313} are essential for the perturbative
calculations of physical observables in high energy processes involving hadrons.
Historically, the development of the factorization theorems as well as the resummation
techniques was based on the analysis of the conventional perturbative QCD Feynman
diagrams. Recently, the soft-collinear effective theory (SCET) was proposed, which
provides a natural framework to deal with the infrared and collinear structure of the QCD
theory. The factorization theorems and the resummation formulas were re-derived in this
framework.

\subsection{The pQCD approach}

The QCD resummation formalism was developed about two decades ago by Dokshitzer, Diakonov
and Troian (DDT) in the double leading logarithm approximation (DLLA)
\cite{hep-ph/0409313}. In order to resum the sub-leading logarithms, transverse momentum
conservation must be imposed. This was achieved by performing Fourier transform to impact
parameter ($b$) space . Based on previous work, Collins, Soper and Sterman showed that
all the large logarithms can be systematically resummed. Their results are often referred
to as the CSS formalism.
Besides the $b$-space formalism, it has been shown that some of the sub-leading
logarithms can also be resummed directly in $q_T$ space (for transverse momentum
distribution) \cite{qtspace}. Their results are consistent with the $b$-space results
up to NNLL level.

After the invention of the resummation formalism, it has been applied to a large variety
of physical processes, including $q_T$ resummation and/or threshold resummation for
vector boson production, Higgs boson production and many other processes. The accuracy of
the calculations have been considerably improved, partly because higher order terms are
included, and partly because of the progress on the non-perturbative parametrization. We
will only review the recent works here.

Most recently, Moch, Vermaseren and Vogt completed the calculation of the three loop
splitting functions \cite{splitting}, which enables one to extract one of the third order
coefficients for the resummation formula. Utilizing this result, two groups independently
worked out the N$^3$LL threshold resummation for Drell-Yan process and Higgs boson
production \cite{nnnll}.

The progresses on the non-perturbative functions are also significant. Landry, Brock,
Nadolsky and Yuan presented a global fit of the non-perturbative parameters based on
their parametrization. They showed that all the available data is in good agreement with
the theoretical predictions, which is strong evidence for the universality of the
non-perturbative function. Kulesza and Stirling \cite{JHEP.0312.056} investigated the
form of the non-perturbative parametrization in both $b$-space and $q_T$-space, and
discussed the theoretical errors in the resummed Higgs $q_T$ distribution arising from
the non-perturbative contribution. They proposed to use $\Upsilon$ production data to
study the non-perturbative contribution in processes with two gluons in the initial
state. Qiu and Zhang \cite{Phys.Rev.D63.114011} proposed an extrapolation method using
conditions of continuity, which Berger and Qiu \cite{qiu} used to present a NNLL
prediction for the Higgs boson $q_T$ distribution.

There are also improvements of the resummation formula in recent years. The original
formula involves process-dependent form factors and coefficient factors. Catani, de
Florian and Grazzini \cite{Nucl.Phys.B596.299} presented a new universal form, in which
the process dependence is embodied in a single perturbative factor. Bozzi, Catani, de
Florian and Grazzini gave a NNLL prediction for the Higgs boson production based on the
above formula. Ji, Ma and Yuan \cite{tmd} proposed a factorization and resummation
formalism in terms of the transverse momentum dependent parton distributions. Berge,
Nadolsky, Olness and Yuan \cite{Phys.Rev.D72.033015} discussed the transverse momentum
resummation with small-$x$ effects taken into account.

The application of the transverse momentum resummation to other processes include: gauge
boson pair production, single stop production, polarized $W$ and $Z$ production at RHIC,
$\Upsilon$ production \cite{Phys.Rev.D57.6934} and single slepton production
\cite{hep-ph/0507331}.

Given the formula for the transverse momentum resummation and the threshold resummation,
one natural question is whether the two effects can be combined. This was achieved as the
so-called joint resummation. The predictions have been worked out for Drell-Yan process,
Higgs boson production as well as top quark pair production \cite{Phys.Rev.D63.114018}.

\subsection{SCET approach}

SCET was proposed in Ref.~\cite{scet} as a systematic framework for the study of
processes involving highly energetic quarks and gluons. In this section we review only a
few key points of SCET and its application to hard processes.

SCET separates the contributions from different energy scales by introducing fields with
well-defined momentum scaling. These modes must reproduce the IR behavior of the full
theory of QCD. We make use of two theories: SCET$_{\rm{I}}$ and SCET$_{\rm{II}}$,
defined in terms of a small scaling parameter $\lambda$. The fields in SCET$_{\rm{I}}$
include: (1) collinear quarks $\xi_n$ and gluons $A_n$ with momenta $p_c \sim
Q(\lambda^2,1,\lambda)$; (2) usoft quarks $q_{us}$ and gluons $A_{us}$ with momenta
$p_{us} \sim Q(\lambda^2,\lambda^2,\lambda^2)$. In SCET$_{\rm{II}}$, the corresponding
fields are: (1) collinear modes with momenta $p_c \sim Q(\lambda^4,1,\lambda^2)$; (2)
soft modes $q_s$, $A_s$ with momenta $p_{s} \sim Q(\lambda^2,\lambda^2,\lambda^2)$. Here
and below we use the light-cone notation $p = (n\cdot p, \bar n\cdot p, p_\perp)$,
defined in terms of light-cone vectors $n$ and $\bar{n}$ satisfying $n^2 = \bar{n}^2 = 0$
and $n \cdot \bar{n} = 2$.

Through integrating out the hard $\sim Q^2$ fluctuations from the full QCD, we get
SCET$_{\rm I}$, and by further integrating out the jets with $\sim Q^2\lambda^2$
fluctuations, SCET$_{\rm{II}}$ can be obtained. This two-step matching can be represented
by $\rm{QCD}\rightarrow \rm{SCET_I}\rightarrow \rm{SCET_{II}}$.

The interactions between the SCET fields are described by the SCET Lagrangians. For the
theory describing usoft and collinear fields, the Lagrangian can be written as~\cite{sii}
\[
  \mathcal{L}_{\mathrm{SCET}} = \mathcal{L}_{\xi\xi} + \mathcal{L}_{cg} +
  \mathcal{L}_{q\xi}.
\]
The Lagrangian can be expanded in $\lambda$. At leading order in $\lambda$,
$\mathcal{L}_{q\xi}$ does not contribute, and the collinear quark Lagrangian is
$$\mathcal{L}_{\xi\xi} = \bar \xi_n \left\{ n\cdot iD_{\rm
    us} + gn\cdot A_n + i\not\!\! D_{\perp c} \frac{1}{\bar n\cdot iD_c} i\not\!\!
  D_{\perp c} \right\} \frac{\not\!\bar n}{2}\xi_n,$$ with $iD^\mu_{\rm us} =
i\partial^\mu + g A^\mu_{\rm us}$. The explicit form of $\mathcal{L}_{cg}$ can be found
in Ref.~\cite{rpi}.

Comparing with the full theory of QCD, an advantage of the SCET is the factorization of
usoft and collinear degrees of freedom at leading order in $\lambda$. Since the usoft
gluons couple to collinear quarks and gluons only through a term $n\cdot A_{us}$,  its
effects can be absorbed into a Wilson line $Y_n[n\cdot A_{us}]$ by a field redefinition:
$$\xi_n = Y_n[n\cdot A_{\rm us}]\xi_n^{(0)},\ \ \ A_n^\mu = Y_n A_n^{(0)\mu} Y^\dagger_n ,$$
$$Y_n[n\cdot A_{\rm us}] \equiv P\exp\left(ig \int_{-\infty}^x ds n\cdot
A_{\rm us}(ns)\right).$$ The new collinear fields $\xi_n^{(0)}$
and $A_n^{(0)}$ do not couple to the usoft gluon field $A_{\rm
us}$, which now appears only through the Wilson line $Y[n\cdot
A_{\rm us}]$. This provides a convenient approach to
SCET$_{\mathrm{II}}$ \cite{iii}.

The applications of SCET to the phenomenology of B decay have been discussed by many
authors \cite{b}\footnote{For further details we refer to the talk of Bauer in this
  proceeding.}. Here we only summarize its applications to the high energy hard
processes. For these processes SCET naturally realizes the proof of factorization. The
matching and running procedure automatically separates the process-dependent Wilson
coefficients and universal quantities in the effective theory, and resums the large
logarithms that may be invalid in the perturbative expansion using the renormalization
group equation directly.

The first investigation was made in Ref.~\cite{sceth}, where the factorization theorem
for various hard processes are proved. Then the enhanced nonperturbative effects in Z
decays as $T\rightarrow 1$ to hadrons was discussed in the framework of SCET \cite{en},
and Ref.~ \cite{t} discussed the resummation for the deep inelastic scattering as
$x\rightarrow 1$. The author of Ref.~\cite{dy} performed the threshold resummation for
Drell-Yan process as $z\rightarrow 1$ in SCET and the transverse momentum resummation for
Drell-Yan process as $Q_T\rightarrow 0$ using SCET was given in Ref.~\cite{qt}.

Furthermore, SCET provides a convenient method to classify and parameterize the
factorizable and nonfactorizable terms beyond the leading order in $\lambda$
\cite{iii,bcdf}. Therefore SCET is a good framework for discussing processes with soft
and collinear particles, and its applications are still developing.

{\em Acknowledgements}: C. S. Li thanks Y. Gao, L. G. Jin, Q. Li, J. J. Liu and L. L.
Yang for their collaborations and many discussions. The authors thank them for their help
with the preparation of this talk, and also thank M. Spira for some useful
suggestions.This work was supported in part by the Natural Sciences Foundation of China.


\begin{thebibliography}{0}
\bibitem{THDM} T. D. Lee, Phys.Rev.D 8 (1973) 1226.
\bibitem{3HDM} S. Weinberg, Phys. Rev. Lett. 37 (1976) 65.
\bibitem{Triplet} H. Georgi and M. Machacek, Nucl.Phys.B 315 (1985)
463; R.S.Chivukula and H. Georgi, Phys.Lett.B 182 (1986) 1981;
P.Galison, Nucl. Phys. B 232 (1984) 26; M. S. Chanpwitz and M.
Golden, Phys. Lett. B 165 (1985) 105.
\bibitem{mssm}
H.~E.~Haber and G.~L.~Kane, Phys. Rep. 117 (1985) 75.
\bibitem{lhiggs}N. Arkani-Hamed et. al.,
Phys. Lett. B 513 (2001) 232; N. Arkani-Hamed et. al., JHEP 0208
(2002) 021, JHEP 0207 (2002) 034.
\bibitem{fath}R. Harnik et. al., Phys. Rev. D70 (2004) 015002.
\bibitem{nggh}A. Diouadi et. al., Phys. Lett. B264 (1991) 440; S.
Dawson, Nucl. Phys. B359 (1991) 283; D. Gradudenz et. al., Phys.
Rev. Lett 70 (1993) 1372; M. Spira et. al., Phys. Lett. B318 (1993)
347; M.Spira et. al., Nucl. Phys. B453 (1995) 17.
\bibitem{nnlo1} R. V. Harlander, Phys. Lett. B492 (2000) 74.
\bibitem{nnlo2} S.Catani et. al., JHEP 0105 (2001) 025; R. V.
Harlander and W. B. Kilgore, Phys. Rev. D64 (2001) 013015.
\bibitem{nnlo3} R. V. Harlander and W. B. Kilgore, Phys. Rev. Lett
88 (2002) 201801; C. Anastasiou and K. Melnikov, Nucl. Phys. B646
(2002) 220; V. Ravindran et. al., Nucl. Phys. B 665 (2003) 325,
Pramana 62 (2004) 683.
\bibitem{ggrr}Z. Bern et. al., Phys. Rev. D66 (2002) 074018.
\bibitem{jetv}D. de. Florian et. al., Phys. Rev. Lett 82 (1999)
5209; S.Catani et. al., JHEP 0201 (2002) 015; V. Ravindran et. al.,
Nucl. Phys. B634 (2002) 247; C. J. Glosser and C. R.Schmidt, JHEP
0212 (2002) 016.
\bibitem{htt}W. Beenakker et. al., Nucl. Phys. B653 (2003) 151,
Phys. Rev. Lett 87 (2001) 201805; S.Dawson et. al., Phys. Rev. D68
(2003) 034022.
\bibitem{hbb}S. Dawson et. al., Phys. Rev. D69
(2003) 074027.
\bibitem{hj1}S. Dittmaier et. al., Phys. Rev. D70
(2004) 074010.
\bibitem{ma}W. Peng et. al., hep-ph/0505086.
\bibitem{gg2}G. P. Gao et. al., Phys. Rev. D71 (2005) 095005.
\bibitem{will}F. Maltoni et. al., Phys. Rev. D67
(2003) 093005.
\bibitem{hj2} J. Campbell et. al., Phys. Rev. D67 (2003) 095002, hep-ph/0405302.
\bibitem{Huang:1998vu}
  C.~S.~Huang and S.~H.~Zhu,
  Phys.\ Rev.\ D {\bf 60}, 075012 (1999)
  [arXiv:hep-ph/9812201].
\bibitem{gbjet}S. Dawson et. al., Phys. Rev. Lett.94 (2005) 031802.
\bibitem{sqcdgb}J. Cao et. al., Phys. Rev. D68 (2003) 075012.
\bibitem{WBF}T. Han et. al., Phys. Rev. Lett. 69 (1992) 3274.
\bibitem{WBFJ}T. Figy and D. Zeppenfeld, Phys. Lett. B591 (2004)
297; T.Figy et. al., Phys. Rev. D68 073005.
\bibitem{eetth}S. Dawson and L. Reina, Phys. Rev. D59 (1999) 054012;
S. Dittmaier et. al., Phys. Lett. B441 (1998) 383.
\bibitem{rrtth}C. Hui et. al., Nucl. Phys. B683 (2004) 196.
\bibitem{massh0}
H.~E.~Haber and R.~Hempfling, Phys. Rev. Lett.66, 1815(1991);
Y.~Okada et al., Prog. Theor. Phys. 85, 1(1991); J.~Ellis et al.,
Phys. Lett. B257, 83(1991); S.~Heinemeyer, hep-ph/0407244.
\bibitem{detect}
A.~Djouadi, Pramana. 62, 191(2004), CERN TH/2003-043,
hep-ph/0303097; M.~Dittmar, talk given at WHEPP 1999, Pramana 55,
151(2000); F.~Gianotti, talk given at the LHC Committee Meeting,
CERN, 5/7/2000.
\bibitem{LHCHiggs}F.~Gianotti, et.al., Eur. Phys. J. C39, 293(2005), CERN-TH/2002-078, hep-ph/0204087;
D.~Denegri et.al., hep-ph/0112045.
\bibitem{gg2h} H.~Georgi et al.,
Phys. Rev. Lett. 40, 692(1978).
\bibitem{susyh} R. V. Harlander and W. B. Kilgore, JHEP 0210 (2002)
017; V. Ravindran et. al., Nucl. Phys. B665 (2003) 325; C.
Anastasiou and K. Melnikov, Phys. Rev. D67 (2003) 037501; B. Field
et. al., Phys. Lett. B551 (2003) 137.
\bibitem{har2} R. V. Harlander and M. Steinhauser, Phys. Rev. D68
(2003) 111701.
\bibitem{hj3} S. Dawson et. al., Phys. Rev. D69 (2004) 074027.
\bibitem{gbth}S. H. Zhu, Phys. Rev. D67 (2003) 075006; T. Plehn,
Phys. Rev. D67 (2003) 014018; E. Berger et. al., Phys. Rev. D {\bf
71} (2005) 115012.
\bibitem{gbthsqcd} G. Gao et. al., Phys. Rev. D66 (2002) 015007.
\bibitem{bbwh}W. Hollik and S. H. Zhu, Phys. Rev. D65 (2002) 075015.
\bibitem{zhaojun}
J.~Zhao, C.~S.~Li and Q.~Li,
  arXiv:hep-ph/0509369, to appear in Phys. Rev. D.
\bibitem{hpair}H. S. Hou et. al., Phys. Rev. D71 (2005) 075014.
\bibitem{hhpair}S. Dawson et al., Phys. Rev. D58 (1998) 115012; T. Plehn et al., Nucl. Phys.
B479 (1996) 46; B351 (1998) 655(E); A. Belyave et al., Phys. Rev.
D60 (1999) 075008; A. A. Barrientos Bendezu and B. A. Kniehl,
Phys. Rev. D64 (2001) 035006.
\bibitem{jin2}L. G. Jin, C. S. Li, Q. Li, J. J.
Liu and R. J. Oakes,
  Phys. Rev. D 71 (2005) 095004.
\bibitem{az} Q. Li, C. S. Li, J. J. Liu, L. G. Jin and C. P. Yuan,
  Phys. Rev. D72 (2005) 034032.
\bibitem{lctth}S.Dittmaier et. al., Phys. Lett. B478 (2000) 24; P. Hafliger and M. Spira, Nucl.Phys. B719
(2005) 35; S. H. Zhu, hep-ph/0212273.
\bibitem{lctbh}B. A. Kniehl et. al., Phys. Rev. D66 (2002) 054016.
\bibitem{Phys.Rev.Lett.74.2905}
  W.~Beenakker et.al.
  Phys.\ Rev.\ Lett.\  {\bf 74}, 2905 (1995);
 ibid,
  Z.\ Phys.\ C {\bf 69}, 163 (1995);
 ibid,
  Nucl.\ Phys.\ B {\bf 492}, 51 (1997).
\bibitem{Nucl.Phys.B515.3}
  W.~Beenakker et.al.,
  Nucl.\ Phys.\ B {\bf 515}, 3 (1998).
\bibitem{berger}
  E.~L.~Berger et.al.,
  Phys.\ Lett.\ B {\bf 459}, 165 (1999)
  [arXiv:hep-ph/9902350];
  ibid.,
  Phys.\ Rev.\ D {\bf 62}, 095014 (2000)
  [arXiv:hep-ph/0005196];
  (E) {\bf 67}, 099901 (2003).
\bibitem{hep-ph/0211145}
  M.~Spira,
  arXiv:hep-ph/0211145.
\bibitem{lgjin}
  L.~G.~Jin,
  C.~S.~Li and J.~J.~Liu,
  Phys.\ Lett.\ B {\bf 561}, 135 (2003);
 ibid., Eur.\ Phys.\ J.\ C {\bf 30}, 77 (2003).
\bibitem{gaugino}
  H.~Baer et.al.,
  Phys.\ Rev.\ D {\bf 57}, 5871 (1998);
  W.~Beenakker et.al.,
  Phys.\ Rev.\ Lett.\  {\bf 83}, 3780 (1999).
\bibitem{PL.B488.359}
  T.~Plehn,
  Phys.\ Lett.\ B {\bf 488}, 359 (2000).
\bibitem{PL.B558.165}
  A.~Alves, O.~Eboli and T.~Plehn,
  Phys.\ Lett.\ B {\bf 558}, 165 (2003).
\bibitem{NP.B660.343}
  D.~Choudhury, S.~Majhi and V.~Ravindran,
  Nucl.\ Phys.\ B {\bf 660}, 343 (2003).
\bibitem{hep-ph/0507331}
  L.~L.~Yang, C.~S.~Li, J.~J.~Liu and Q.~Li,
  Phys.\ Rev.\ D {\bf 72}, 074026(2005).
\bibitem{hep-ph/9809259}
  M.~Kramer,
  Nucl.\ Phys.\ Proc.\ Suppl.\  {\bf 74}, 80 (1999).
\bibitem{ee}
  A.~Arhrib et.al.,
  Phys.\ Rev.\ D {\bf 52}, 1404 (1995).
  H.~Eberl et.al.,
  Nucl.\ Phys.\ B {\bf 472}, 481 (1996).
\bibitem{NP.B515.15}
  C.~H.~Chang et.al.,
  Nucl.\ Phys.\ B {\bf 515}, 15 (1998).
\bibitem{qqg}
  A.~Brandenburg, M.~Maniatis and M.~M.~Weber,
  arXiv:hep-ph/0207278.
\bibitem{topdecay1}
A.~Czarnecki, Phys. Lett. B {\bf 252}, 467 (1990);
C.~S.~Li, R.~J.~Oakes and T.~C.~Yuan, Phys. Rev. D {\bf 43}, 3759 (1991).
\bibitem{two}
M.~Slusarczyk,  hep-ph/0404249.
\bibitem{pol}
M.~Fischer et.al., 
Phys. Lett. B {\bf 451}, 406 (1999); ibid, Phys. Rev. D {\bf 64}, 017301 (2001); D {\bf
  65}, 054036 (2002); W.Bernreuther et.al., 
Phys. Lett. B {\bf 582}, 32 (2004); A. Brandenburg et.al., 
Phys. Lett. B {\bf 539}, 235(2002).
\bibitem{dist}
A.~Brandenburg, M. Maniatis, Phys.Lett. B {\bf 545}, 139 (2002).
\bibitem{susde}
J.~J.~Liu, C.~S.~Li, L.~L.~Yang, L.~G.~Jin., Phys. Lett. B {\bf 599}, 99 (2004).
\bibitem{t1}
M.~C.~Smith et.al., 
Phys. Rev. D {\bf 54}, 6696 (1996); C.~S. Li, R.~J.~Oakes, J.~M.~Yang, H.~Y.~Zhou, Phys.
Rev. D {\bf 57}, 2009 (1998).
\bibitem{t2}
G. Bordes et.al., 
Nucl. Phys. B {\bf 435}, 23 (1995);
T. Stelzer et.al., 
Phys. Rev. D {\bf 56}, 5919 (1997).
\bibitem{t3}
S.~H.~Zhu, Phys. Lett. B {\bf 524}, 283 (2002); (E) {\bf 537}, 351 (2002).
\bibitem{diff}
B.~W.~Harris et.al., 
Phys. Rev. D {\bf 66}, 054024 (2002).
\bibitem{diy}
Q.~H.~Cao, R.~Schwienhorst and C.-P.~Yuan, Phys. Rev. D {\bf 71}, 054023 (2005);
Q.~H.~Cao and C.-P.~Yuan Phys. Rev. D {\bf 71}, 054022 (2005);
Q.~H.~Cao, R.~Schwienhorst, J.~A.~Benitez, R.~Brock and C.-P.~Yuan, hep-ph/0504230;
J.~Campbell, R.~K.~Ellis and F.~Tramontano, Phys. Rev. D {\bf 70}, 094012(2004).
\bibitem{l1}
C.~S.~Li, X.~M.~Zhang, S.~H.~Zhu, Phys. Rev D {\bf 60}, 077702 (1999).
\bibitem{hera}
A.~Belyaev and N.~Kidonakis, Phys. Rev. D {\bf 65}, 037501 (2002).
\bibitem{np}
J.~J.~Liu, C.~S.~Li, L.~L.~Yang, L.~G.~Jin, Nucl. Phys. B {\bf 705}, 3 (2005);
ibid., Mod. Phys. Lett. A19, 317 (2004).
\bibitem{qcdfc}
N.~Kidonakis et.al.,
JHEP 0312 (2003) 004;
J.~J.~Liu, C.~S.~Li, L.~L.~Yang, L.~G.~Jin, Phys. Rev. D {\bf 72}, 074018 (2005).
\bibitem{eg}
J.~H.~Kuhn, C.~Sturm, P.~Uwer, Eur. Phys. J. C {\bf 30} 169 (2003).
\bibitem{nason}
P.~Nason et.al., 
Nucl.Phys. B303, (1988) 607; W.Beenakker et.al., 
Phys.Rev. D40, (1989) 54.
\bibitem{ptt}
P.Nason et.al., 
Nucl.Phys. B327 (1989)49; W. Beenakker et. al., 
 Nucl.Phys. B351 (1991) 507;
M.L.Mangano et.al., 
Nucl.Phys. B373 (1992)295; S. Frixione et.al., 
Phys. Lett. B351 (1995)555;
S.Catani et. al., 
Phys. Lett. B378 (1996)329; Nucl. Phys. B478 (1996)273; E.L.Berger
et.al., 
Phys. Rev. D57 (1998)253;
N. Kidonakis et. al. 
Nucl. Phys. B505 (1997) 321; R.Bonciani et. al., 
Nucl. Phys. B529 424(1998).
\bibitem{top1}
N. Kidonakis, Int. J. Mod. Phys. A15 (2000) 1245; Mod. Phys. Lett.
A19(2004)405; Int. J. Mod. Phys. A20 (2005) 3726; N. Kidonakis,
Phys. Rev. D64 (2001)014009; Int. J. Mod. Phys. A16 Suppl. 1A, 363
(2001); N.
Kidonakis et.al., 
Phys. Rev. D64 (2001)114001; Phys. Rev. D67 (2003)074037; Nucl.
Phys. A715 (2003)549 N. Kidonakis et. al. 
Phys. Rev. 68 (2003) 114014; Eur. Phys. J. C33 (2004) s466.
\bibitem{spin2}
 W.Bernreuther et. al., 
 Phys. Lett. B509
(2001) 53; Phys. Rev. Lett. 87 (2001) 242002; Int.J.Mod.Phys. A18
(2003) 1357.
\bibitem{spin3}
A. Brandenburg, Z.G. Si  Phys.Lett. B615 (2005) 68
\bibitem{zhu}
 S. Godfrey, S.H. Zhu, hep-ph/0412261.
\bibitem{tt3}
A.Brandenburg, S.Dittmaier, P.Uwer, S.Weinzierl, hep-ph/0408137
\bibitem{lin}
C. Macesanu,  Phys. Rev. D65 (2002) 074036; C.Macesanu, L.H. Orr,
Int. J. Mod. Phys. A16S1A (2001) 369; Phys. Rev. D65 (2002) 014004.
\bibitem{cslin}
C.H.Chang, C.S. Li, R.J.Oakes, J.M.Yang, Phys. Rev. D51, 2125
(1995); A.Djouadi, M.Drees, and h.Konig, Phys. Rev. D48, 3081
(1993)
\bibitem{csli}
H.Wang, C.S. Li, H.Y. Zhou, Y.P. Kuang, Phys. Rev. D54, 4374
(1996).
\bibitem{csli2}
C.S.Li et.al., 
Phys. Rev. D52, 5014-5017 (1995), Erratum-ibid. D53, 4112 (1996);
C.S. Li et. al., 
 Phys. Lett. B379 (1996)
135; C.S. Li et.al., 
Phys. Lett. B424 (1998) 76.
\bibitem{csli3}
H.Y.Zhou, C.S.Li, Phys.Rev.D55:4421-4429,1997.
\bibitem{spin1}
H. X. Liu, C.S. Li, Z. J. Xiao Phys. Lett. B458 (1999) 393; J.
Kodaira, T.Nasuno, S. Parke, Phys.Rev. D59 (1999) 014023 ;
A.Brandenburg, M. Maniatis,  Phys. Lett. B558 (2003) 79.
\bibitem{hep-ph/0409313}
  J.~C.~Collins et. al., 
  Adv.\ Ser.\ Direct.\ High Energy Phys.\  {\bf 5}, 1 (1988);
  Y.~L.~Dokshitzer et. al., 
  Phys.\ Lett.\ B {\bf 79}, 269 (1978);
  Y.~L.~Dokshitzer et. al., 
  Phys.\ Rept.\  {\bf 58}, 269 (1980);
  G.~Parisi et.al. 
  Nucl.\ Phys.\ B {\bf 154}, 427 (1979);
  G.~Curci et.al., 
  Nucl.\ Phys.\ B {\bf 159}, 451 (1979);
  J.~C.~Collins et.al., 
  Nucl.\ Phys.\ B {\bf 193}, 381 (1981);
  [Erratum-ibid.\ B {\bf 213}, 545 (1983)];
 ibid,
  Nucl.\ Phys.\ B {\bf 197}, 446 (1982);
  J.~C.~Collins et.al., 
  Nucl.\ Phys.\ B {\bf 250}, 199 (1985).
\bibitem{qtspace}
  R.~K.~Ellis, D.~A.~Ross and S.~Veseli,
  Nucl.\ Phys.\ B {\bf 503}, 309 (1997).
  R.~K.~Ellis et.al., 
  Nucl.\ Phys.\ B {\bf 511}, 649 (1998);
  A.~Kulesza et.al., 
  Nucl.\ Phys.\ B {\bf 555}, 279 (1999).
\bibitem{splitting}
  S.~Moch et.al., 
  Nucl.\ Phys.\ B {\bf 688}, 101 (2004);
  A.~Vogt et. al., 
  Nucl.\ Phys.\ B {\bf 691}, 129 (2004).
\bibitem{nnnll}
  S.~Moch et. al., 
  arXiv:hep-ph/0508265;
  E.~Laenen et. al., 
  arXiv:hep-ph/0508284.
\bibitem{JHEP.0312.056}
  A.~Kulesza et. al., 
  JHEP {\bf 0312}, 056 (2003).
\bibitem{Phys.Rev.D63.114011}
  J.~w.~Qiu et. al., 
  Phys.\ Rev.\ D {\bf 63}, 114011 (2001).
\bibitem{qiu}
  E.~L.~Berger et. al., 
  Phys.\ Rev.\ Lett.\  {\bf 91}, 222003 (2003);
  E.~L.~Berger et. al., 
  Phys.\ Rev.\ D {\bf 67}, 034026 (2003).
\bibitem{Nucl.Phys.B596.299}
   S.~Catani et. al., 
  Nucl.\ Phys.\ B {\bf 596}, 299 (2001).
\bibitem{tmd}
  X.~d.~Ji et. al., 
  Phys.\ Lett.\ B {\bf 597}, 299 (2004);
 ibid, Phys.\ Rev.\ D {\bf 71}, 034005 (2005).
\bibitem{Phys.Rev.D72.033015}
  S.~Berge et. al., 
  Phys.\ Rev.\ D {\bf 72}, 033015 (2005).
\bibitem{Phys.Rev.D57.6934}
  C.~Balazs et. al., 
  Phys.\ Rev.\ D {\bf 57}, 6934 (1998);
  T.~Plehn,
  Phys.\ Lett.\ B {\bf 488}, 359 (2000);
  A.~Weber,
  Nucl.\ Phys.\ B {\bf 403}, 545 (1993);
  P.~M.~Nadolsky et. al., 
  Nucl.\ Phys.\ B {\bf 666}, 31 (2003);
  E.~L.~Berger et. al., 
  Phys.\ Rev.\ D {\bf 71}, 034007 (2005).
\bibitem{Phys.Rev.D63.114018}
  E.~Laenen et. al., 
  Phys.\ Rev.\ D {\bf 63}, 114018 (2001);
  A.~Kulesza et. al., 
  Phys.\ Rev.\ D {\bf 66}, 014011 (2002);
  A.~Kulesza et. al., 
  Phys.\ Rev.\ D {\bf 69}, 014012 (2004);
  A.~Banfi et. al., 
  Phys.\ Rev.\ D {\bf 71}, 034003 (2005).
\bibitem{scet}
C.~W.~Bauer, S.~Fleming, and M.~E.~Luke, Phys. Rev. D63,
014006(2001); C.~W.~Bauer, S.~Fleming, D.~Pirjol, and
I.~W.~Stewart, Phys. Rev. D63, 114020(2001); C.~W.~Bauer and
I.~W.~Stewart, Phys. Lett. B516, 134(2001).
\bibitem{sii}
C.~W.~Bauer, D.~Pirjol, and I.~W.~Stewart, Phys. Rev. D65,
054022(2002).
\bibitem{rpi}
J.~Chay and C.~Kim, Phys. Rev. D65, 114016(2002).
\bibitem{iii}
C.~W.~Bauer, D.~Pirjol, and I.~W.~Stewart,
 Phys. Rev. D68, 034021(2003).
\bibitem{b}
J. Chay and C.Kim, Phys. Rev. D68 (2003) 071502; ibid, Phys. Rev.
D68 (2003) 034013; S.W.Bosch, R.J.Hill, B.O.Lange, and M.Neubert,
Phys. Rev. D67, 094014 (2003) C.W. Bauer and A.V. Manohar, Phys.
Rev. D70, 034024 (2004); M. Beneke and T. Feldmann, Nucl. Phys.
B685, 249 (2004).
\bibitem{sceth}
C.W. Bauer, S. Fleming, D.Pirjol, I.Z. Rothstein, and I.W.Stewart,
Phys. Rev. D66, 014017 (2002).
\bibitem{en}
C.W. Bauer, A.V.Manohar, and M.B.Wise, Phys. Rev. Lett. 91,
122001(2003); ibid, Phys. Rev. D70, 034014 (2004).
\bibitem{t}
A.V.Manohar, Phys.Rev.D 68, 114019(2003).
\bibitem{dy}
A.Idilbi and X.D.Ji, Phys. Rev. D72 (2005) 054016.
\bibitem{qt}
Y.Gao, C.S.Li, and J.J.Liu, hep-ph/0501229; A.Idilbi, X.D.Ji, and
F.Yuan, Phys. Lett. B625 (2005) 253.
\bibitem{bcdf}
M.~Beneke, A.~P.~Chapovsky, M.~Diehl, and T.~Feldmann, Nucl. Phys.
B643, 431(2002).
\end{thebibliography}
\end{document}